*Article*

# Targeted Polariton Flow Through Tailored Photonic Defects

Elena Rozas [1,*] , Yannik Brune [1], Ken West [2], Kirk W. Baldwin [2], Loren N. Pfeiffer [2], Jonathan Beaumariage [3], Hassan Alnatah [3], David W. Snoke [3] and Marc Aßmann [1]

1   Department of Physics, TU Dortmund University, Dortmund 44227, Germany; yannik.brune@tu-dortmund.de (Y.B.); marc.assmann@tu-dortmund.de (M.A.)
2   Department of Electrical Engineering, Princeton University, Princeton, NJ 08544, USA; kwb@princeton.edu (K.W.B.); loren@princeton.edu (L.N.P)
3   Department of Physics  Astronomy, University of Pittsburgh, Pittsburgh, PA 15260, USA; jcb130@pitt.edu (J.B.); haa108@pitt.edu (H.A.); snoke@hey.com (D.W.S.)
*   Correspondence: elena.rozas@tu-dortmund.de (E.R.)

**Abstract:** In non-Hermitian open quantum systems, such as polariton condensates, local tailoring of gains and losses opens up an interesting possibility to realize functional optical elements. Here, we demonstrate that deliberately introducing losses via a photonic defect, realized by reducing the quality factor of a DBR mirror locally within an ultrahigh-quality microcavity, may be utilized to create directed polariton currents towards the defect. We discuss the role of polariton-polariton interactions in the process and how to tailor the effective decay time of a polariton condensate via coupling to the defect. Our results highlight the far-reaching potential of non-Hermitian physics in polaritonics.

## 1. Introduction

While photonic defects can often be the source of unwanted losses in optical structures, they can also enable novel optical functionalities by confining and guiding light at specific frequencies [1–5]. These defects, which may arise from disruptions in the periodic structure of photonic crystals or from localized deformations that can degrade the quality factor, can significantly alter the optical properties of the material [6–9]. For this reason, when used deliberately, photonic defects can provide precise control over light, especially in confined systems, facilitating the creation of localized modes that can trap electromagnetic waves, much like electrons in solid-state systems.

Generally, photonic defects can be classified based on their nature and the mechanisms by which they alter light, including vacancy defects, line defects, or more complex structures [10–12]. For instance, photonic quantum well structures create confined quantized states that mimic electronic tunneling effects in semiconductor quantum wells [13,14]. These states illustrate the diversity of optical capabilities that defects offer in photonic structures, from controlling spontaneous emission rates to engineering the spatial mode distributions within a cavity. Despite their potential, research on tailoring defect states in optical microcavities is still limited, revealing a significant gap in the field of non-Hermitian open quantum systems, such as polariton condensates.

In optical microcavities, where the cavity size is comparable to the photon wavelength, light is confined through resonant recirculation [15–17]. This confinement modifies the density of optical modes compared to free space, allowing precise control over light-matter interactions and therefore, over polaritons, which emerge from the strong coupling between excitons and photons within the microcavity [18–20]. Given that polaritons, as non-equilibrium systems, exhibit both gain and loss mechanisms, they offer a unique platform for exploring the effects of photonic defects. As a result, the presence of defects can uncover rich phenomena like variations in the effective decay rates of the condensate. Additionally, such controlled dissipation can also enable the creation of directed polariton flows [21], potentially leading to the development of innovative optical elements, such as switches[22–25], gates[26,27] or routers [28,29] for photonic circuits.

In this work, we explore the impact of a localized photonic defect embedded in an ultrahigh-quality microcavity on an optically trapped polariton condensate. By varying the distance between the polariton condensate and the defect, we observe significant changes in the polariton flow and occupation levels. We analyze separately the momentum distribution for both the defect and the polariton system, providing a comprehensive understanding of their mutual influence. In addition, we demonstrate how these types of defects result in an increase of polariton-polariton interactions and cause a redistribution of the condensate's energy modes when both systems are in close proximity. Our findings emphasize the critical role of photonic defects in polaritonic systems and highlight their potential for engineering polariton currents and controlling condensate dynamics [30,31].

## 2. Materials and Methods

In this work, we employ an ultrahigh Q factor $3\lambda/2$ optical microcavity, which incorporates 12 GaAs quantum wells (each with a thickness of 7 nm) positioned between two distributed Bragg reflector (DBR) mirrors. The quantum wells are grouped into three sets of four, with each group positioned at one of the three antinodes of the cavity. The DBRs, made from alternating layers of AlAs and $Al_{0.2}Ga_{0.8}As$, are composed of 32 layers on the top mirror and 40 layers on the bottom mirror [32]. This configuration leads to a long polariton lifetime of approximately 200 ps [33]. The measurements were performed in a region of the sample with a cavity-exciton detuning of $\delta_{C-X} \approx -0.8$ meV.

Throughout the experiments, the sample was kept at 10 K using a cold-finger flow cryostat. The sample was excited nonresonantly at the first Bragg minimum of the cavity, 1.746 eV, with an ultra narrow linewidth cw Ti:Sapphire laser. A spatial light modulator (SLM) was used to modify the phase front and shape the intensity distribution of the excitation beam into a 10 µm diameter ring, creating an optical trap for polariton condensate formation. This configuration was used to minimize the decoherence effects arising from carrier-polariton interactions [32]. To achieve polariton condensation within the optical trap, a minimum power threshold of $P_{th}$ = 86 mW was required. The condensate was created in the vicinity of a photonic defect with a FWHM of 5 µm. The defect was created through a controlled process using a high-power laser. This process introduces localized losses into the system, leading to a localized reduction in the quality factor in that specific region of the microcavity. A microscope objective with 20x magnification and a numerical aperture of 0.26 was employed for both exciting the sample and collecting its photoluminescence (PL). The PL polarization was restricted to the polarization component parallel to the main axis of the cavity, which also aligns with the polarization direction of the polariton condensate. Finally, a spatial filter was integrated into the setup to single out emission either from the ring excitation region or the defect in the surrounding area.

## 3. Results and discussion

Photonic defects in microcavities significantly influence the polariton dynamics by locally altering the optical properties of the cavity. These defects can modify not only the flow of polaritons, but also the cavity-exciton coupling as well as the overall losses within the system. Therefore, understanding how polaritons interact with these defects is crucial for controlling the interplay between polariton currents and their photonic environment. For this reason, in the following subsections, we first explore the influence of a photonic defect on polariton currents and the corresponding mode distribution as the distance between the two systems is varied. We then explore the unique phenomena that arise when the defect and polariton condensate fully overlap.

### 3.1. Polariton Flow Near Photonic Defects

To illustrate these effects, we first examine how the polariton flow from the excitation beam contributes to the defect population. To do so, we excite the sample above the condensation threshold, at 1.4 $P_{th}$, and then collect the PL intensity from the defect for distances up to 200 µm, where our field of view reaches its limit. These results are depicted in Figure 1(a). The PL intensity of the defect exhibits an exponential decay with increasing distance, resulting in a total intensity reduction of 3 orders of magnitude as the defect reaches 200 μm of separation from the trapped condensate. These findings demonstrate that the defect can effectively serve as a gauge for the polariton flow, revealing the extent of polariton propagation within the microcavity. The observed long-range behavior of the defect population suggests that polaritons travel even greater distances than the measured ones, likely due to the long polariton lifetime in the microcavity used in this study.

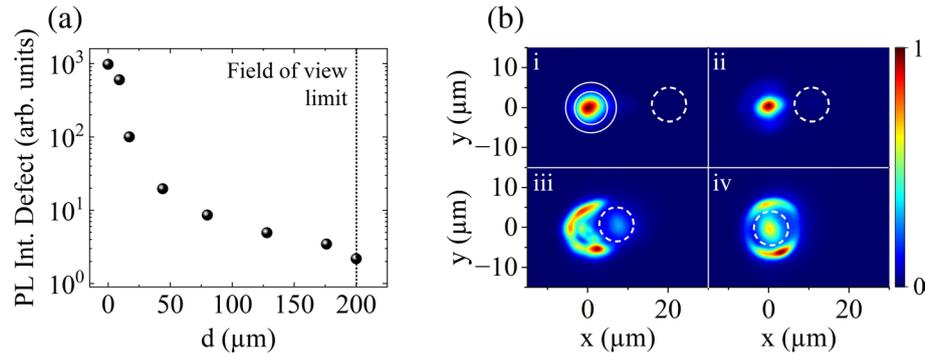

**Figure 1.** (a) PL intensity of the defect measured at different distances (d) from the polariton condensate. (b) Polariton condensate PL at different distances: (i) 20 μm, (ii) 10 μm, (iii) 7 μm and (iv) 0 μm. The solid lines in panel (i) depict the ring-shaped excitation profile with a diameter of 10 μm. The dashed lines indicate the location of the defect. All measurements were performed at P = 1.4 $P_{th}$.

Despite the optical trapping of the polariton condensate at the center of the ring, we still observe a polariton flow propagating toward the defect. This behavior can be attributed to scattering processes and a small portion of polaritons being ejected away from the excitation area. As a result, these polaritons can reach the defect site, leading to the observed PL emission from the defect even when it is positioned far from the condensate. After establishing the polariton flow's long-range behavior, we next focus on the detailed effects observed at shorter distances. Particularly interesting effects emerge at distances below 20 μm, where significant changes in the polariton condensate are observed as the defect approaches the optical trap. Figure 1(b) presents the polariton condensate PL in real-space at distances ranging from d = 20 to 0 μm. As shown in panel (i), the condensate is generated at the center of a ring-shaped optical trap, indicated by solid lines, while the defect, indicated by dashed lines, is located in the surrounding area. At 20 μm, the polariton condensate exhibits the highest occupation level, while the PL intensity of the defect is several orders of magnitude lower, rendering it not visible in the image. As the distance to the defect decreases, a redistribution of the polariton occupation becomes evident. For the case of d = 10 μm, a soft reduction in the condensate density is observed. However, when the defect approaches closer to the ring trap, at 7 μm, the condensate is no longer present at the center of the trap, revealing the presence of the circular barrier. At the same time, the PL of the defect becomes more pronounced. Finally, at 0 μm, when the excitation beam and the defect fully overlap, a substantial portion of the polariton flow is redirected toward the defect site, resulting in the highest defect emission intensity compared to other distances.

While the analysis of the real-space polariton distribution can provide direct insights into the interaction between polariton currents and defects, it is also essential to understand how the defect influences the momentum distribution of the system. For this reason, we use the spatial filter to selectively chose the emission from one of the two systems and remove contributions from the other one, providing a more comprehensive picture of the momentum-space characteristics. This analysis is illustrated in Figure 2, where panel (a) shows the real-space emission of the system below condensation threshold at d = 20 μm. The filtered areas for the ring-shaped potential and the defect are depicted with a cyan and a yellow square, respectively. Once the area of interest has been selected, Fourier spectroscopy is performed to obtain the corresponding momentum-space distribution. Panel (b) and (e) display the LPB band at low pump power for the filtered ring-shaped area and defect area, respectively. In both panels, the white solid line marks the energy value of 1.6026 eV, which corresponds to the bottom of the parabolic dispersion band of the defect state. Interestingly, the dispersion measured at the ring area shows slightly higher energy values. This small decrease in energy caused by the local decrease in the quality factor of the microcavity facilitates the existence of a polariton flow towards lower energies, resulting in a gain in the occupation level as the defect gets closer to the excitation beam.

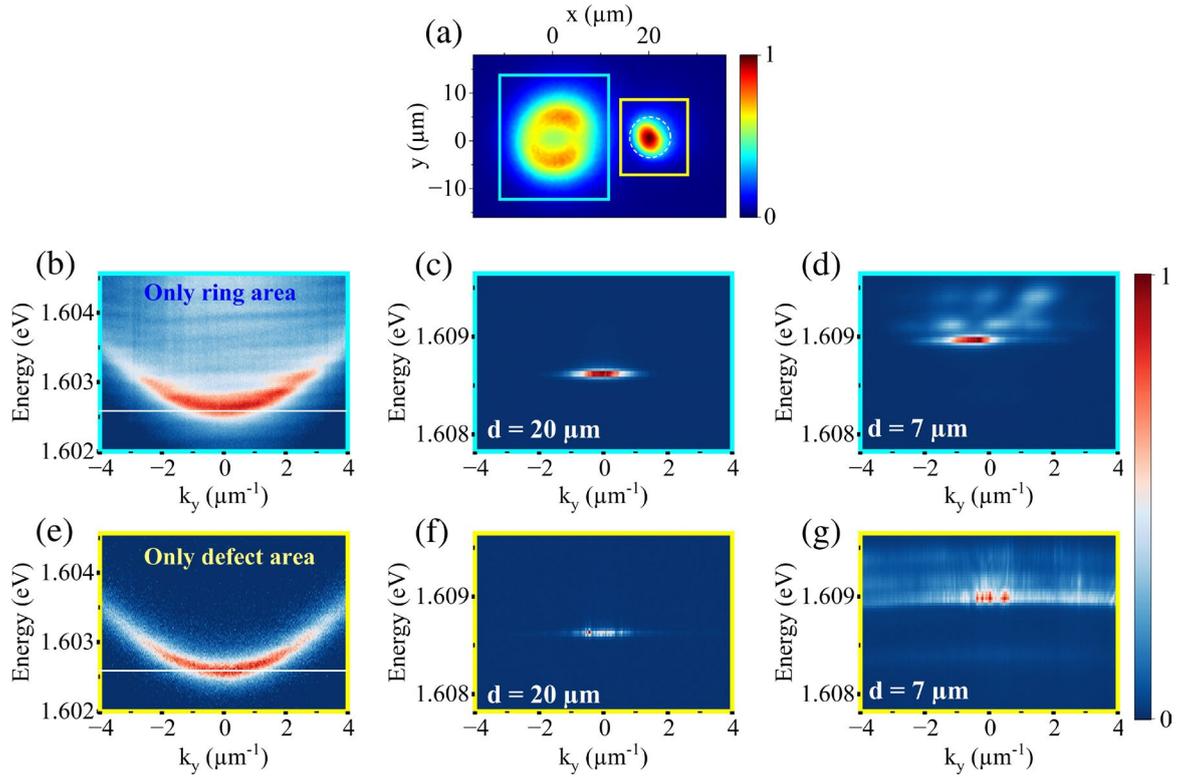

**Figure 2.** (a) Real space PL of the system at $0.5P_{th}$ and d = 20 µm. The cyan square indicates the ring-shaped excitation area, while the yellow square marks the defect area. (b) Momentum space image obtained by collecting emission only from the ring-shaped area at $0.5P_{th}$. The white solid line marks the ground energy of the defect state, 1.6026 eV. (c)-(d) Momentum space image of the condensate formed within the ring area at $1.4P_{th}$ for distances d = 20 µm and 7 µm, respectively. (e) Momentum space image of the defect obtained by collecting emission only from the defect area at $0.5P_{th}$. (f)-(g) Momentum space image of the defect at $1.4P_{th}$ for distances d = 20 µm and 7 µm, respectively. All momentum distributions have been measured along $k_y$-axis.

Furthermore, we observe a small broadening of the LPB band within the photonic defect. In particular, at $0.5\,P_{th}$, the spectral linewidth increases from $\gamma_r = 0.30$ meV in the ring area to $\gamma_d = 0.39$ meV in the defect area. Although precise estimates of polariton lifetimes are challenging to obtain, this broadening indicates the presence of altered polariton dynamics. This evidence points to increased cavity losses, and thus a reduction in the radiative polariton lifetime within the defect.

When the pump power is increased to $1.4P_{th}$, and nonlinear polariton-polariton scattering becomes efficient, the polariton condensate forms inside the optical trap. At d = 20 µm, see panel (c), the condensate exhibits a momentum distribution centered at $k_y = 0$. A large blueshift of the emission of 5-6 meV can be observed, which results from polariton-polariton interactions and interactions with dark excitons with k-vectors beyond the light cone [32,34]. As the distance between the condensate and the defect is further reduced, the momentum distribution experiences a strong change. At 7 µm, panel (d), the defect overlaps with the optical trap, deforming the ring potential barrier. As a consequence, the polariton condensate strongly interacts with the polariton population in the defect, resulting in a further increase of the polariton-polariton interactions and therefore, an increase in the blueshift of the emission. In addition, the emission exhibits simultaneously a small shift in momentum, with a new center at $k_y = -0.5\,\mu m^{-1}$, and a multimode emission, with the onset of localized states at different energies. However, the corresponding real space image for d = 7 μm in Figure 1(a)iii, shows that the peak emission is not centered within the trap. This leads to the possibility that while the polariton flow generated by the excitation beam is feeding the defect, the trapped condensate is generated in the ring barrier. At the same time, the momentum distribution of the defect under the same conditions, see Figure 2(f)-(g), shows an emission centered at $k_y = 0$ for all distances and a blueshift comparable to that of the condensate. This continuous increase in blueshift, driven by polariton-polariton interactions, reflects a rising polariton occupation in the defect, which is consistent with our initial observations presented in Figure 1.

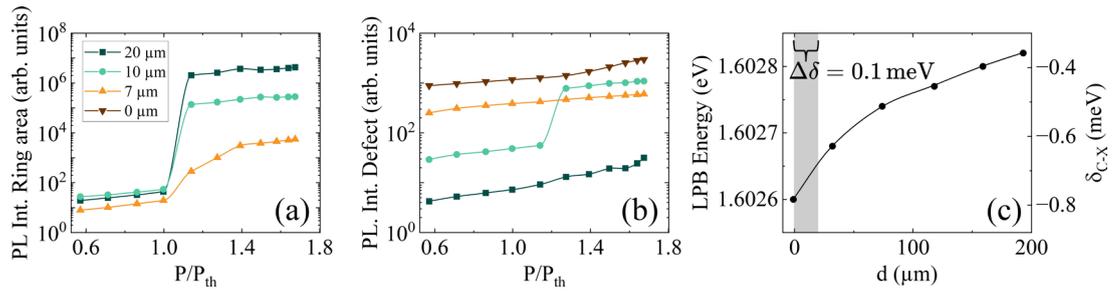

**Figure 3.** (a) PL intensity of the ring-shaped area as a function of pump power, measured at various distances from the defect. To avoid duplication of data within the same figure, the case d = 0 μm is only shown in panel (b). (b) PL intensity of the defect area as a function of pump power, obtained for the same distances as in (a). (c) Cavity-exciton detuning dependence with distance. The gray-shaded area represents the 20 μm region where all the measurements were performed, corresponding to a detuning variation of 0.1 meV.

To understand the interaction between the two systems in detail, we carefully analyze the PL intensity of these emissions for a wide range of pump powers. Figure 3(a) illustrates the PL intensity when only the ring area is spatially filtered. At larger distances from the defect, where interactions between both systems are minimal, we observe strong nonlinearities during condensate formation, reflected by an increase in the PL intensity of 5 orders of magnitude. However, as the defect approaches, the condensate intensity decays, indicating that the defect is indeed draining the system. Consequently, at 7 μm, the condensation threshold is no longer well defined, and the observed strong nonlinearities fade. On the other hand, the PL intensity when the defect is measured is shown in Figure 3(b). Here, a different trend emerges. At 20 μm, the defect's occupancy is relatively low compared to the other distances. As the distance decreases to 10 μm, there is a clear rise in intensity, and a condensation threshold becomes apparent within our range of measured powers. It is noteworthy that the nonlinear increase in the defect PL intensity occurs at a power 15% higher compared to the threshold of the trapped condensate, indicating that the condensate forms in the ring trap before this increase in the defect is observed. As the defect moves into the ring barrier, its PL intensity continues to rise, reaching its maximum value when the defect is centered at the ring. These results demonstrate that the reduced quality factor near the defect induces a polariton flow away from the condensate and toward the defect.

It is also important to note that the sample has a wedge in the cavity that modulates the cavity-exciton detuning, $\delta_{C-X}$, which affects the excitonic content of polaritons and may influence how polaritons propagate along the sample. Figure 3(c) displays the LPB energy for different distances within our field of view, and its corresponding detuning. When the excitation beam is located 200 μm away from the defect, polaritons exhibit a detuning of ∼ −0.4 meV. As they get closer to the defect, the LPB energy decreases, evidencing a decrease in their excitonic content. Therefore, this evolution potentially facilitates the polariton flow toward the defect due to the energy gradient and a reduction in their effective mass. However, in small areas such as the studied region of 20 μm, see highlighted gray area in Figure 3(c), the total change in detuning is 0.1 meV, which is not expected to significantly alter the polariton properties.

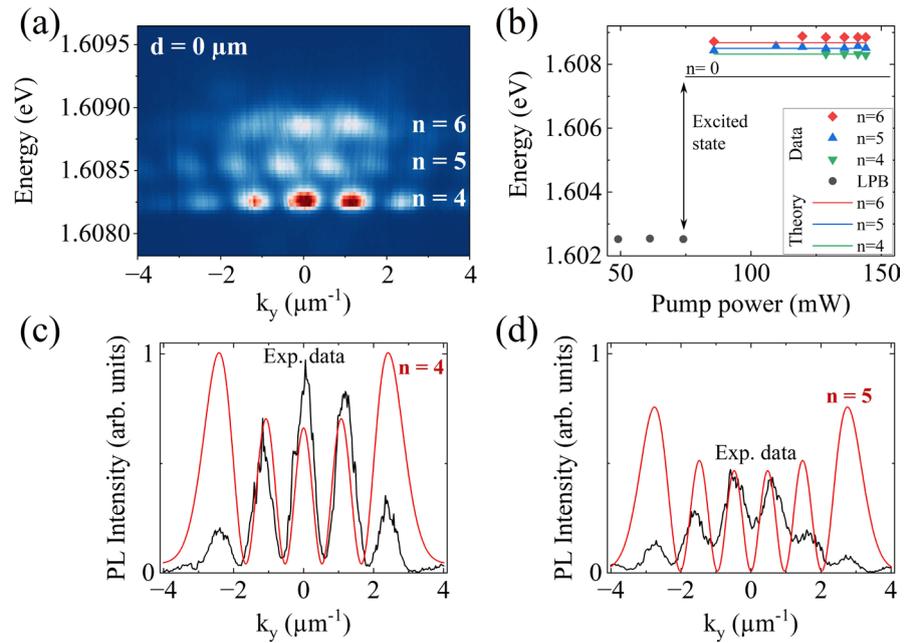

**Figure 4.** (a) Angle-resolved emission for the case d = 0 μm and $P$ = 140 mW, showing a QHO distribution with three distinct energy levels. (b) Energy of the observed states as a function of pump power. Experimental data is represented by dots and theoretical approximations of each energy level by solid lines. The black line indicates the $n$ = 0 level of the excited state, which corresponds to the ground state of the polariton condensate. (c)-(d) Theoretical fits for the eigenstates $n$ = 4 and $n$ = 5, respectively. In both panels, the experimental data is extracted from panel (a).

### 3.2. Generation of a Quantum Harmonic Oscillator Potential

In contrast to most of the studied distances in this work, at d = 0 μm, we observe additional distinct phenomena that differ from our previous observations. When the populated photonic defect overlaps with optically trapped polaritons, localized energy states with a discrete momentum distribution become visible. The first traces of this effect were observed at d = 7 μm, Figure 2(d), however, only at d = 0 μm the localization is fully observable. Figure 4(a) illustrates the emission spectrum of the system for the latter case. Three distinct energy levels with localized momentum distributions indicate the presence of a Quantum Harmonic Oscillator (QHO) potential. Therefore, we initially identify these states as the eigenstates n = 4, n = 5 and n = 6. This case provides a unique opportunity to examine the conditions under which these states appear and how different energy levels are populated. For this reason, in panel (b) we analyze the energy of the different observed states as a function of pump power. We compare the experimental data, represented by dots, with the theoretical approximations, represented by solid lines. The calculated energy levels 1.60832 eV (n = 4) and 1.60850 eV (n = 5), fit our data with very good agreement, confirming the appearance of a QHO potential in the system. However, the level presumed to correspond to n = 6 does not match the theoretical value of 1.60868 eV, leaving its exact nature uncertain.

To further explore these results, we extracted the intensity profiles along $k_y$ for the n = 4 and 5 states from panel (a). In Figure 4(c)-(d) we present theoretical fits for these experimental data, demonstrating that the position of the PL intensity maxima along momentum align with the expected QHO distribution. Nevertheless, the PL intensity distribution exhibits a more Gaussian profile than is typically expected for a QHO. This can be attributed to the fact that at 0 µm, the defect is located at the center of the optical trap, constantly redirecting the polariton flow. This central position can increase significantly the polariton occupation at $k_y = 0$, which may contribute to the observed Gaussian shape and hinder the expected intensity distribution for a QHO.

## 4. Conclusions

In conclusion, we have demonstrated the significant impact of localized photonic defects on the behavior of polariton condensates within ultrahigh-quality microcavities. By strategically introducing a defect that increases the losses in the structure, we have shown that it is possible to guide polariton currents toward the defect and modify the interplay between gain and loss mechanisms inherent to polariton systems. Adjusting the proximity of the condensate to the defect leads to significant changes in the polariton flow and occupancy levels, underscoring the critical role of defects and offering a method to modify polariton-polariton interactions. Moreover, our findings reveal that the presence of photonic imperfections transforms the condensate emission from typical single-mode to more complex multimode emission, indicating a rich dynamic interaction between defects and polaritons. Notably, when the defect fully overlaps with optically trapped polaritons, we observe distinct phenomena not seen at larger distances. Localized energy states with discrete momentum distributions emerge, resembling the eigenstates of a Quantum Harmonic Oscillator potential. Our analysis confirms the presence of these QHO-like states, particularly for the levels $n = 4$ and 5. However, further investigation is required to fully understand the nature of higher excited states. The results presented here show the potential of photonic defects to enable precise control over light-matter interactions, paving the way for new applications in signal processing.

**Author Contributions:** Conceptualization, M.A. and E.R.; methodology, E.R. and Y.B.; formal analysis, E.R.; investigation, E.R., Y.B. and M.A; sample design J.B., H.A. and D.W.S.; sample fabrication K.W., K.W.B. and L.N.P.; writing—original draft preparation, E.R.; writing—review and editing, E.R. and M.A; visualization, E.R.; supervision, M.A.; All authors have read and agreed to the published version of the manuscript.

**Funding:** This research was funded by the QuantERA II Programme, which received funding from the EU H2020 research and innovation programme under GA No. 101017733, and by the Deutsche Forschungsgemeinschaft (DFG) within the projects under GA No. 231447078 and 532767301. The Princeton University acknowledges financial support from the Gordon and Betty Moore Foundation's EPiQS Initiative, Grant GBMF9615.01 to L. Pfeiffer. The Pittsburgh team acknowledges financial support from the National Science Foundation, grant DMR-2306977.

**Institutional Review Board Statement:** Not applicable.

**Data Availability Statement:** The data supporting the conclusions of this article will be made available on request by the corresponding authors.

**Conflicts of Interest:** The authors declare no conflicts of interest.

## Abbreviations

The following abbreviations are used in this manuscript:

| | |
|---|---|
| DBR | Distributed Bragg reflector |
| SLM | Spatial Light Modulator |
| PL | Photoluminescence |
| $P_{th}$ | Power Threshold |
| LPB | Lower Polariton Branch |
| QHO | Quantum Harmonic Oscillator |